\documentclass[nofootinbib,apsspec,showpacs,twocolumn,groupedaddress,lengthcheck,preprintnumbers]{revtex4}
\usepackage{amsmath}
\usepackage{amssymb}
\usepackage{graphics}
\usepackage[dvips]{graphicx}
\usepackage{graphicx}
\usepackage[usenames]{color}

\newcommand \beq{\begin{eqnarray}}
\newcommand \eeq{\end{eqnarray}}
\def\simge{\mathrel{%
       \rlap{\raise 0.511ex \hbox{$>$}}{\lower 0.511ex \hbox{$\sim$}}}}
\def\simle{\mathrel{
       \rlap{\raise 0.511ex \hbox{$<$}}{\lower 0.511ex \hbox{$\sim$}}}}

\definecolor{purple}{rgb}{0.6, 0.2, 0.8}

\usepackage[colorlinks=true,linktocpage=true,linkcolor=blue,citecolor=blue]{hyperref}

\begin{document}
\title{Virtual photon polarization in ultrarelativistic heavy-ion collisions }
\author{Gordon Baym,$^{a,b}$ Tetsuo Hatsuda,$^{b,c}$, and Michael Strickland$^d$}
\affiliation{\mbox{$^a$Department of Physics, University of Illinois, 1110
  W. Green Street, Urbana, IL 61801-3080, United States} \\
\mbox{$^b$iTHES Research Group and iTHEMS Program, RIKEN, Wako, Saitama 351-0198, Japan} \\
\mbox{$^c$Nishina Center, RIKEN, Wako, Saitama 351-0198, Japan} \\
\mbox{$^d$Department of Physics, Kent State University, Kent, OH 44242, United States}
}


\begin{abstract}
The polarization of direct photons
produced in an ultrarelativistic heavy-ion collision reflects the 
momentum anisotropy of the quark-gluon plasma created in the collision.   
This paper presents a general framework, based on the photon spectral functions in the plasma, for analyzing the angular distribution and thus the polarization of dileptons  in terms of the plasma momentum anisotropies. 
The rates of dilepton production depend, in general, on four independent spectral functions, corresponding to two  
transverse polarizations, one longitudinal polarization, and -- in plasmas in which the momentum anisotropy is not invariant under parity in the local rest frame of the matter -- a new spectral function, $\rho_n$, related to the anisotropy direction in the collision.  The momentum anisotropy appears in the difference of the two transverse spectral functions, as well as in $\rho_n$.   As an illustration, we delineate the spectral functions for dilepton pairs produced in the lowest order Drell-Yan process of quark-antiquark annihilation to a virtual photon.
\end{abstract}

\preprint{RIKEN-QHP-251}

\maketitle

\section{Introduction}

     Direct photons, both real and virtual, are an important probe of the dynamics of ultrarelativistic heavy-ion collisions.  An average temperature of 
the quark-gluon plasma (QGP) formed in high-energy collisions has been extracted from the transverse momentum spectrum of direct photons in the range $q_{\rm T}\sim $ 1-3 GeV  \cite{PHENIX, STAR, ALICE}.   Theoretically, measurements of the  photon polarization
through the angular distribution of  dileptons ($l^+\l^-$) have been proposed to provide information on the early stages of collisions, before the onset of thermalization \cite{Hoyer,Bratkovskaya,Shuryak}.     While relativistic hydrodynamics provides
a  successful spacetime description of the later stages of the collision dynamics
and associated hadronic and leptonic  observables \cite{QM2015_proc},
important questions concerning the early dynamics, such as the degree of thermalization as well as isotropization of the QGP, have not been answered either experimentally or theoretically.    Recently, Ref.~\cite{BH} proposed using the polarization of direct photons as a measure of the gluon anisotropy in collisions.  While measuring direct photon polarization, involving external conversion to dilepton pairs, is very difficult experimentally, a more promising approach to is measure polarization of 
virtual photons, through the angular distribution of dileptons produced via internal conversion.  

\begin{figure}[h]
\begin{center}
\includegraphics[width=0.65\linewidth]{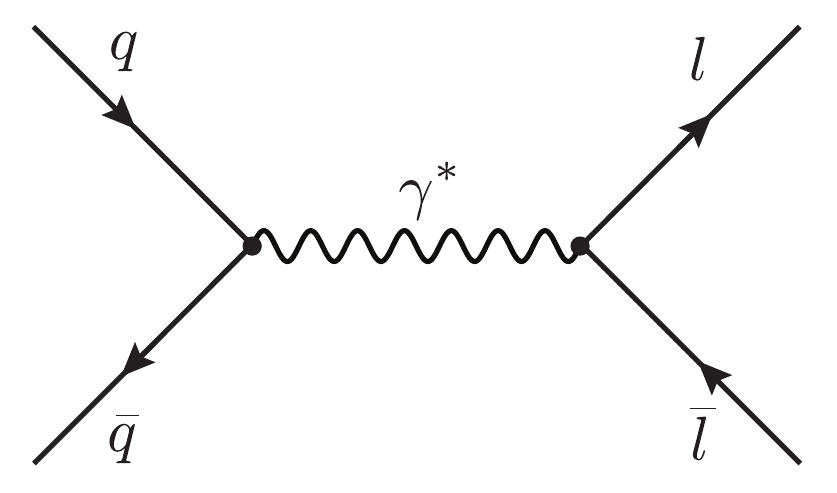}
\vspace{-5mm}
\end{center}
\caption{Lowest order Drell-Yan production of a lepton pair. }
\label{dy}
\end{figure}

   The lowest order mechanism to produce dilepton pairs is the Drell-Yan process,  Fig.~\ref{dy},  in which a quark and an antiquark annihilate to a virtual photon.    The dilepton cross section
$d\sigma/d\Omega$ can be parametrized as  $\propto
 1+\lambda \cos^2 \theta + \mu \sin 2\theta \cos \phi + (\nu/2) \sin^2 \theta \cos 2 \phi$,  
  where $\theta$ and $\phi$ are the polar and azimuthal angles of one of the dileptons 
  in the dilepton rest frame measured in the Collins-Soper reference frame  \cite{Collins-Soper,Lam-Tung}.  High-energy $p\bar{p}$ and $pp$ collisions at the Tevatron and the
  LHC have confirmed the leading-order prediction $\lambda\simeq 1$, $\mu = \nu = 0$, 
 for $q_{\rm T} \simle $ 5 GeV and  invariant dilepton mass $M_{l^+l^-} \simeq M_Z$ (see, e.g.,~\cite{Peng} and references therein).  
The dilepton angular distribution in In-In collisions has been measured by the NA60 experiment \cite{NA60} at the CERN-SPS 
in the primary kinematical range $0.4 < M_{l^+ l^-} < $  0.9 GeV, in which production of dileptons by hadronic 
sources such as $\pi^+\pi^-$ annihilation dominates Drell-Yan dileptons; the results are consistent with $\lambda=\mu=\nu=0$.

   On the other hand,  in ultrarelativistic heavy-ion collisions we expect 
   the momentum-space anisotropy of the QGP 
   to be detected  most readily in dilepton production in the range  1 GeV $< q_{\rm T} <$ 3 GeV 
   and $M_{l^+ l^-} \simle  0.3 $ GeV
    where the excess of direct photons is seen experimentally \cite{PHENIX}.        
   In this paper, we present a general framework for analyzing the angular distribution of dileptons emitted from a QGP, which can be used to extract quark and gluon momentum anisotropies in the collision. 
The basic starting point is the photon polarization tensor and associated spectral function;  we analyze the spectral function 
in terms of four vectors, two transverse polarization vectors,
a longitudinal polarization vector, and a vector specifying the  
momentum anisotropy.   Such a decomposition combined with the leptonic tensor describing the conversion of a virtual photon into a dilepton pair leads to a  general formula relating the momentum anisotropy of the plasma and
the dilepton angular distribution.  As an illustration, we apply the formalism to virtual photon emission through the leading-order  Drell-Yan process in a plasma with anisotropic distributions of the Romatschke-Strickland form \cite{RS}.\footnote{In the course of writing this paper we became aware of the work of Friman and collaborators \cite{bengt} which does not include the anisotropic terms $\rho^T_1-\rho^T_2$ and $\rho_n$, but otherwise arrives at results in agreement with those given here; the
approach of these two treatments of the problems are complementary and will be discussed in a future joint publication of the two groups.}

\section{Dilepton production}

  Quite generally, the production rate, $R$, of a dilepton pair is proportional to the spectral function, $\rho_{\mu\nu}$, of the in-medium photon polarization or self-energy operator, for momentum $\vec q$ and energy $q^0$,
\beq
\Pi_{\mu\nu}(\vec q\,,z) = e^2 \int_{-\infty}^{\infty}\frac{dq^0}{2\pi} \frac{\rho_{\mu\nu}(\vec q\,,q^0)}{z-q^0} \, ,
\eeq
times the squared matrix element $L^{\mu\nu}$ (L for leptons) for a virtual photon of 4-momentum $q$ to produce a lepton of 4-momentum $p$ and mass $m$ and an antilepton of 4-momentum $p'$, averaged over the spins of the leptons.
 Explicitly, 
\beq
  \frac{dR_{l^+  l^-}}{d^3\bar{p} d^3\bar{p}'}   = \frac{\alpha^2}{4\pi^4 Q^4}  \  \rho_{\mu\nu} (q)  L^{\mu\nu}(p,p') \, ,
  \label{rate}
\eeq
with the leptonic tensor,
\beq
L_{\mu\nu} (q,s) =2  \left( q_\mu q_\nu -g_{\mu\nu} Q^2 -s_\mu s_\nu \right) ,
\eeq
where  $d^3\bar{p} \equiv  d^3p/2E_p$, $d^3\bar{p}' \equiv  d^3p^\prime /2E_{p'}$, $q=p+p'$,
$s=p-p'$, $Q^2 \equiv q^\mu q_\mu > 0$, and $Q^2+s^2=4m ^2$ with $m$ the lepton mass.

\begin{figure}[h]
\begin{center}
\includegraphics[width=0.65\linewidth]{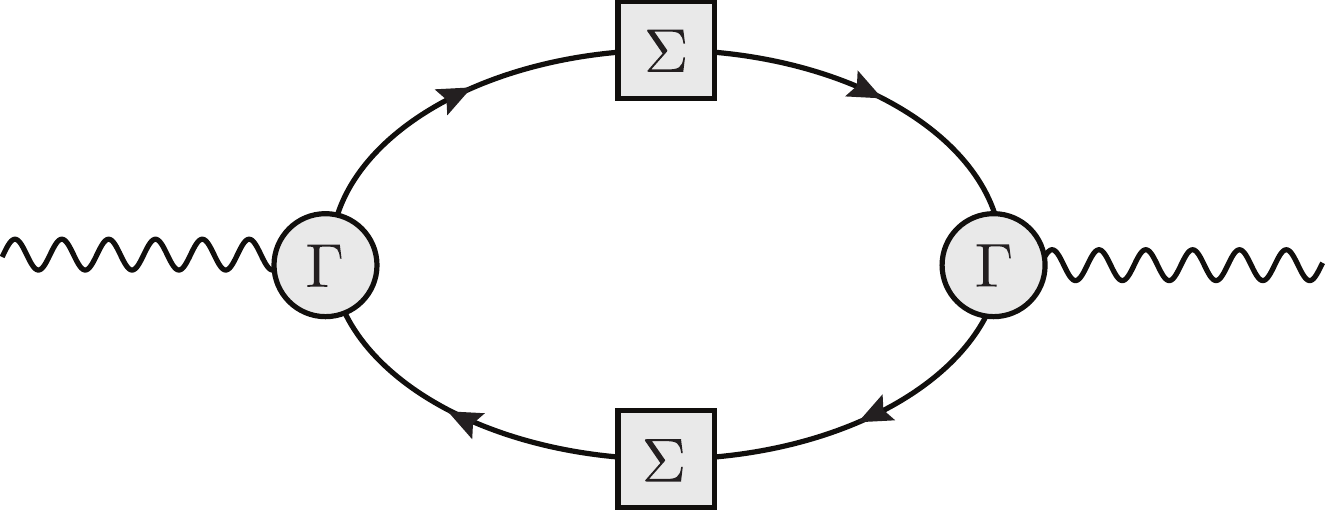} 
\end{center}
\vspace{-5mm}
\caption{Photon polarization tensor with hard thermal loop corrections to the quark lines and vertices \cite{Carrington:2007gt}.
}
\label{bubbles}
\end{figure}

    The spectral function  $\rho_{\mu\nu} (q)$ is related to the cut, or imaginary part, of the photon polarization operator, illustrated
in Fig~\ref{bubbles}.  Its explicit form  in the kinematical regime $q^0 \gg T \gg \sqrt{Q^2}$ has been previously evaluated using hard thermal loop effective theory for the isotropic quark-gluon plasma   \cite{Carrington:2007gt,Laine:2015iia}.   The heart of the problem in this paper is to determine the structure of $\rho_{\mu\nu}(q)$,  to see how the anisotropy of the gluon and quark distributions is reflected in the  final orientation of the dilepton pair.

\section{Structure of $\rho_{\mu\nu}$}

\begin{figure}[h]
\vspace{-36pt}
\begin{center}
\includegraphics[width=1.1\linewidth]{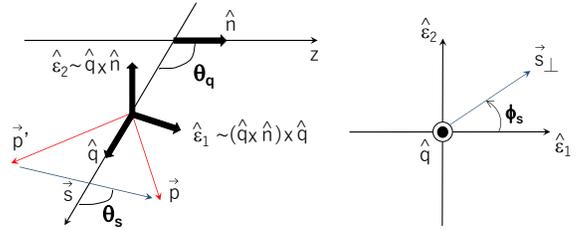} 
\end{center}
\vspace{-8mm}
\caption{(Left) Virtual photon polarization vectors $\hat \varepsilon_i$ and the  relative spatial momentum 
$\vec s$ between the lepton and antilepton. (Right) The relative lepton momentum in the plane transverse to
the virtual photon momentum $\vec{q}$. 
 }
\label{pol-vector}
\end{figure}

   In a heavy-ion collision volume,  the initial gluon and quark distributions are anisotropic in momentum space with a single preferred axis $\hat n$, which we assume to be along the beam direction \cite{Strickland:2013uga}   (we do not consider at this point possible multiple anisotropy axes).   
We define the four vector $n^\mu$ to have space component $\hat n$ in the local rest frame of the matter and time component, $n^0=0$, 
 \beq
    n^\mu = (0, \hat n) \, ,
\eeq        
  so that $n^2=-1$.  We also define, in the local rest frame, the two transverse polarization vectors 
\beq  
  \varepsilon_i^\mu = (0, \hat \varepsilon_i) \, , 
 \eeq 
where 
$\hat\varepsilon_1 \equiv (\vec q \times \hat n) \times \vec q/|\vec q \times \hat n|$ and 
 $\hat \varepsilon_2 \equiv \vec q \times \hat n/|\vec q\times \hat n|$.  These polarization vectors are illustrated in the
 left panel of Fig.~\ref{pol-vector}.  In addition, we define the longitudinal polarization vector 
\beq
   \varepsilon_{\rm L}^\mu \equiv \frac1{\sqrt{Q^2}} (|\vec q\,|,q^0\hat q) =  (|\vec {\tilde q}\,|,\tilde q^0\hat q) \, .
\eeq
where we write $\tilde q^\mu = q^\mu/\sqrt{Q^2}$.
Note that $\varepsilon_1^2 = \varepsilon_2^2 = \varepsilon_{\rm L}^2 =  -1$.  The three polarization vectors are individually orthogonal to $q^\mu$: $(q \varepsilon) = 0$ (where
 $(ab)$ denotes the four vector product of $a$ and $b$) and together with $q^\mu$ form an orthogonal basis obeying;
\beq
    g^{\mu\nu} = \tilde q^\mu \tilde q^\nu - \varepsilon_1^\mu \varepsilon_1^\nu - \varepsilon_2^\mu\varepsilon_2^\nu 
    -\varepsilon_{\rm L}^\mu\varepsilon_{\rm L}^\nu \, .
    \label{g}
\eeq

    Thus the photon spectral function, $\rho_{\mu\nu}$, is a sum of terms of the form
\beq
  a \varepsilon_{\rm L}^\mu\varepsilon_{\rm L}^\nu + b \varepsilon_1^\mu\varepsilon_1^\nu 
 +c ( \varepsilon_1^\mu\varepsilon_{\rm L}^\nu +\varepsilon_{\rm L}^\mu\varepsilon_1^\nu) + d  \varepsilon_2^\mu\varepsilon_2^\nu \, . 
  \label{abcd}
\eeq  
There are no terms proportional to $q^\mu q^\nu$ for $Q^2\ne 0$, since $q$ is a zero eigenvector of $\rho$;  in addition $(n \varepsilon_2)=0$, so there are no $\varepsilon_2\varepsilon_{\rm L}$ terms by symmetry.   
Since $\vec n$ can be written as the linear superposition, $\vec n = \cos \theta_q \hat{q} + \sin \theta_q \vec \varepsilon_2$,  with
$\theta_q$ being the angle between $\hat n$ and $\hat q$, and $n^0=0 $, we obtain
\beq
        n^\mu =\cos \theta_q (\tilde q^0 \varepsilon_{\rm L}^\mu  - |\vec {\tilde q}\,| \tilde q^\mu)      + \sin \theta_q \,\varepsilon_1^\mu \, .
    \label{ntheta}
\eeq
We also introduce the four-vector   ${\cal N}^\mu$ with the property  $({\cal N} q)=0$ as, 
\beq
  {\cal N}^\mu \equiv \tilde q^0\, \cos \theta_q\, \varepsilon_{\rm L}^\mu + \sin\theta_q\,\varepsilon_1^\mu 
   = n^{\mu} - (\tilde{q}n) \tilde{q}^{\mu} ,
\label{eq:defN}
\eeq
  Thus  the 
 $\varepsilon_1^\mu  \varepsilon_{\rm L}^\nu + \varepsilon_{\rm L}^\mu \varepsilon_1^\nu$ term can be eliminated in favor of ${\cal N}^\mu {\cal N}^\nu$, $\varepsilon_1^\mu\varepsilon_1^\nu$, and $\varepsilon_{\rm L}^\mu\varepsilon_{\rm L}^\nu$.  
 In addition,  ${\cal N}^2 = -(1+ (n\tilde q)^2)$.  
  The latter term, plus the explicit $\varepsilon_{\rm L}^\mu\varepsilon_{\rm L}^\nu$ term in Eq.~(\ref{abcd}), can be eliminated using Eq.~(\ref{g}).  
 The photon spectral function, again with the help of Eq.~(\ref{g}), assumes the general form
 \beq
 \rho^{\mu\nu} &=& 
 \varepsilon_{\rm L}^\mu\varepsilon_{\rm L}^\nu\rho^{\rm L} 
 +\varepsilon_1^\mu\varepsilon_1^\nu\rho^{\rm T}_1 
 +\varepsilon_2^\mu\varepsilon_2^\nu \rho^{\rm T}_2 
 + {\cal N}^\mu {\cal N}^\nu  \rho_n    \label{impilong} \\
 &= & -(g^{\mu\nu} -\tilde q^\mu \tilde q^\nu)\rho^{\rm L} 
  +\varepsilon_1^\mu\varepsilon_1^\nu(\rho^{\rm T}_1-\rho^{\rm L}) \nonumber \\ 
  &&+\varepsilon_2^\mu\varepsilon_2^\nu (\rho^{\rm T}_2-\rho^{\rm L}) + {\cal N}^\mu {\cal N}^\nu  \rho_n \, .
   \label{impi}
\eeq  
The momentum-space anisotropy of the system leads to the extra $ \rho_n$ term, as well as a difference of $\rho_1^{\rm T}$ and $\rho_2^{\rm T}$.   

  The terms $\rho_n$, $\rho^{\rm T}_{2}$ can be extracted directly  from Eqs.~(\ref{impilong}) and (\ref{eq:defN}) as 
\beq
   \varepsilon_{1\,\mu} \rho^{\mu\nu}\varepsilon_{{\rm L}\, \nu} = - \tilde q\,^0 \cos \theta_q \, \sin \theta_q \, \rho_n \, ,
   \label{n}
 \eeq
and 
\beq
   \varepsilon_{2\,\mu} \rho^{\mu\nu}\varepsilon_{2\, \nu} =  \rho^{\rm T}_2 \, ,
   \label{t2}
\eeq
while $\rho^{\rm L}$ is found from 
\beq
     \varepsilon_{{\rm L}\,\mu} \rho^{\mu\nu}\varepsilon_{{\rm L}\, \nu} = \rho^{\rm L} +(\tilde q\,^0)^2 \cos^2\theta_q \, \rho_n \, .
   \label{rhoL}
\eeq
Using Eq.~(\ref{n}),   we find $\rho_1^{\rm T}$ from the trace condition,
\beq
     \rho^\mu_\mu =  - (\rho^{\rm L} + \rho^{\rm T}_1 + \rho^{\rm T}_2 + (1+ (n\tilde q)^2)\rho_n)  \, .
\label{eq:trace-rho}
\eeq

   When the particle distribution functions are even under parity, so that $\vec n$ enters only as a special axis, not a special direction,
the extra $\rho_n$ term must vanish.  To see this we note that when the distribution functions are parity invariant, both parity and the
transformation $\hat n\to -\hat n$ are independent symmetries, meaning that a parity transformation keeping $\vec n$ fixed is also a
symmetry.  But under such a transformation $\vec\varepsilon_{\rm L}$ transforms as a vector, while $\vec \varepsilon_1$ transforms as a pseudovector; thus the mixing of the two directions in $\rho_{ij}$, the source of $\rho_n$, cannot occur.    For collisions of two identical nuclei, there should not be a special direction in the local rest frame of the matter.   Below, when we write down the Drell-Yan rate in the medium in such a situation, we will see explicitly how this argument is realized.   However, for asymmetric collisions,
one expects a non-zero $\rho_n$ term in the photon spectral function.   In the following, we keep the $\rho_n$ term in the general discussions.

  The various $\rho$ depend separately on the local $q^0$, $q_\perp$, and $\vec q\cdot \hat n$, where $q_\perp$ is the magnitude of the component of $\vec q$ orthogonal to $\hat n$.    Or expressed covariantly, they depend  on $Q^2$, 
  $(q u)$, as well as on $(q n)$,  where $u_\mu$ is the 4-velocity of the local rest frame.  Note that  
  $(n u) \equiv 0$.
       
    We look now at the eigenvalue structure of  photon spectral function.   In the local rest frame, in an isotropic system or for $\vec q$ along $\hat n$, the $\rho_n$ term vanishes and $\rho^{\rm T}_1=\rho^{\rm T}_2$, while for $\vec q \perp \hat n$ the eigenvectors of $\rho$ are the $\varepsilon_i$, with the two eigenvalues, $\rho^{\rm T}_1$ and $\rho^{\rm T}_2$.    More generally, in the local rest frame, the eigenvectors of $ \rho^{\mu\nu}$  are $q^\mu$ with eigenvalue 0, $\varepsilon_2^\mu$ with eigenvalue $ \rho_2^{\rm T}$,
and two  orthogonal linear combinations of $\varepsilon_1^\mu$ and $\varepsilon_{\rm L}^\mu$ whose spatial components lie in the $(\vec q, \hat n)$ plane. In the isotropic limit, the eigenvector $\varepsilon_{\rm L}$ has eigenvalue $ \rho_{\rm L}$, and $\varepsilon_1$ has eigenvalue $ \rho_{\rm T}$.  In contrast, when the system is anisotropic, the two eigenfunctions describe propagation in a birefringent medium with a mixing of the longitudinal (L) and transverse (T) polarizations.   Furthermore,  as $\vec q\to 0$,  $ \rho^{\rm T}_i -  \rho^{\rm L}$ must vanish as $\vec q\,\,^2$, and thus, $ \rho^{\rm T}_i = \rho^{\rm L}$ for $i = 1,2$.

\section{Emission rate of dileptons and photons}

   To calculate the production rates of dilepton pairs we first note that quite generally, $(sq)=0$, so that
 \beq
 \frac{1}{2} \rho^{\mu\nu}L_{\mu\nu} &=& - (Q^2 \rho^\mu_\mu + s_\mu  \rho^{\mu\nu} s_\nu ) \\
  &= & Q^2\left(\rho^{\rm T}_1+\rho^{\rm T}_2 + \rho_n\right) + 4m^2\rho^{\rm L} \nonumber\\ 
& & -s_1^2(\rho^{\rm T}_1-\rho^{\rm L}) - s_2^2(\rho^{\rm T}_2 -\rho^{\rm L}) \nonumber \\
& & + ((qn)^2- (sn)^2) \rho_n \, , 
\label{saniso}
\eeq 
where we use the identity $Q^2+s^2 = 4 m^2$, with $m$ being the lepton mass, and we define $s_i \equiv (s\varepsilon_i)$ ($i=1,2$) to be the components of $\vec s$ transverse to $\vec q$ in the local rest frame:  $\vec s_\perp = s_1 \vec\varepsilon_1 + s_2\vec \varepsilon_2$.

Equation (\ref{saniso}) gives the dilepton production rate in terms of the projections of $s$ along the two 
transverse polarizations and $n$.
The $s_1^2$ and $s_2^2$ terms  contain the anisotropy produced by transverse virtual photons, while from Eq.~(\ref{ntheta}), we see that the $s_z ^2$ term arises from the mixing of longitudinal and  transverse ($\vec \varepsilon_1$) virtual photons.    As noted above, for symmetric collisions with parity invariance in the local matter rest frame, $\rho_n$ should vanish so then the final term Eq.~(\ref{saniso}) is absent. 
To bring out the anisotropic terms, we write $s_1 =|{\vec s}_\perp| \cos \phi_s $, and 
$s_2 =  |{\vec s}_\perp| \sin\phi_s$;  the squared matrix elements (\ref{saniso}) become
\beq
\frac{1}{2} \rho^{\mu\nu}L_{\mu\nu}  
&= &  2 Q^2  \bar{\rho}^{\rm T}  + \left(s_\perp^2 + 4m^2  \right)   \rho^{\rm L}  \nonumber \\
& & + \left( Q^2 + (qn)^2 - (sn)^2\right)  \rho_n \nonumber \\  
& & -  |\vec{s}_\perp|^2 \left(   \bar{\rho}^{\rm T} +  \delta{\rho}^{\rm T} \cos 2 \phi_s \right) ,
\label{saniso2}
\eeq 
where $\bar{\rho}^{\rm T} \equiv ( \rho_1^{\rm T}  + \rho_2^{\rm T} )/2$ and  
$\delta{\rho}^{\rm T} \equiv ( \rho_1^{\rm T}  -  \rho_2^{\rm T} )/2$. The $\cos 2\phi_s$  as well as the $\rho_n$ terms are anisotropic.
  For massless dileptons in the absence of $\rho_n$, the right side of
Eq.~(\ref{saniso2}) becomes $2 Q^2\bar{\rho}^{\rm T} +  |\vec{s}_\perp |^2 (\rho^{\rm L} -  \bar{\rho}^{\rm T}  -\delta{\rho}^{\rm T} \cos 2 \phi_s)$.   With $\theta_s$ the angle between
$\vec q$ and $\vec s$, we see that this expression
 is of the form $\propto 1 + \lambda_s \cos^2\theta_s + \mu_s \sin 2 \theta_s \cos\phi_s + (\nu_s/2) \sin^2\theta _s\cos 2 \phi_s$, with 
\beq
    \lambda_s = \frac{\bar{\rho}^{\rm T} -\rho^{\rm L}}{ \bar{\rho}^{\rm T}(1- 2 s_0^2/\vec s\,^2) + \rho^{\rm L}}, \nonumber\\
    \nu_s = \frac{-2\delta{\rho}^{\rm T}}{ \bar{\rho}^{\rm T}(1- 2 s_0^2/\vec s\,^2) + \rho^{\rm L}} \, , 
    \eeq
and $\mu_s = 0$.   We note the similarity to the angular distribution fitted in the NA60 analysis \cite{NA60}, where the angles are defined in the Collins-Soper frame;  as noted above, NA60 finds when averaging over all lab directions of the virtual photons, that $\lambda$, $\mu$, and $\nu$ are consistent with zero.

   With the Jacobian from the variables $p$ and $p'$ to $Q$, $ \vec{s}_\perp$,  rapidity $y$ and  $\vec{q}_{_{\rm T}}$, 
\beq
   d^3\bar p \, d^3 \bar{p}' = \frac{1}{2}
    \frac{dQ^2 \,  dy \, d^2q_{_{\rm T}}\, d^2s_\perp}{\sqrt{Q^2(Q^2 - s_\perp^2 - 4m^2)}}\ ,
\eeq 
we finally obtain the dilepton emission rate 
\beq
& &  \frac{dR_{l^+  l^-}}{dQ^2 d^2s_\perp   dy  d^2q_{_{\rm T}}} =   
\frac{\alpha^2}{4 \pi^4 Q^4} \frac{\rho^{\mu\nu}L_{\mu\nu}/2}{\sqrt{Q^2(Q^2 - s_\perp^2 - 4m^2)}} \, , \nonumber\\  
\label{rate2}
\eeq
where $ \rho^{\mu\nu}L_{\mu\nu}/2$ is given by Eq.~(\ref{saniso2}).
As seen in Fig.~\ref{pol-vector},
 the components $(sn)$ and  $(qn)$ are not independent; their dependence on the experimental variables, $Q^2$,  $s_\perp$,  $y$, and  $q_{_{\rm T}}$ is algebraic (but too complicated to quote here). 
  
  It is instructive to connect the present formalism for virtual photons to the calculation of the rate for real photons, $Q^2=0$ including possible polarization, as considered by  \cite{Schenke:2006yp,Bhattacharya:2015ada} and \cite{BH}.
To do so, we rewrite
Eq.~(\ref{impi}) as
 \beq
 \rho^{\mu\nu} &= & -(g^{\mu\nu}Q^2 -q^\mu q^\nu)\frac{ \rho^{\rm L}}{Q^2}  +\varepsilon_1^\mu\varepsilon_1^\nu( \rho^{\rm T}_1- \rho^{\rm L}) \nonumber \\ 
  &&+\varepsilon_2^\mu\varepsilon_2^\nu ( \rho^{\rm T}_2- \rho^{\rm L}) + (Q^2{\cal N}^\mu )( Q^2 {\cal N}^\nu ) \frac{ \rho_n}{Q^4}\, ,    \label{impi0}
\eeq 
from which we see that as $Q^2\to0$, $\rho_{\rm L}$ vanishes as  $Q^2$ and $\rho_n$ vanishes as $Q^4$.
   Thus the rate to produce a real photon with polarization $\varepsilon^\mu$ is 
\beq
\frac{dR_{\gamma}}{d^3 \bar{q}} &=&
  \frac{\alpha}{2\pi^2} \varepsilon^*_{\mu} \rho^{\mu \nu} \varepsilon_{\nu}, \nonumber\\
 &&= \frac{\alpha}{2\pi^2} \left( \bar\rho^{\rm T} +( |(\varepsilon\varepsilon_1)|^2 - |(\varepsilon\varepsilon_2)|^2)\delta\rho^{\rm T} \right) \nonumber\\
 &&= \frac{\alpha}{2\pi^2} \left( \bar\rho^{\rm T} + \delta{\rho}^{\rm T} \cos 2 \phi_\varepsilon \right),
\eeq
 where $(\varepsilon\varepsilon_1) \equiv -\cos\phi_\varepsilon$, $(\varepsilon\varepsilon_2) \equiv -\sin\phi_\varepsilon$, and
$d^3\bar{q}= d^3 q / 2| \vec{q}|$. 
The anisotropy for real photons arises entirely from the difference, $\delta\rho^{\rm T}$, 
of $\rho^{\rm T}_1$ and $\rho^{\rm T}_2$: the spectral function $\rho_n$ does not enter.

\section{Drell-Yan process in the medium}

 To give a specific illustration of the present formalism we focus on the leading-order
  Drell-Yan production of dilepton pairs where
  the squared matrix element for a quark and antiquark to produce a virtual photon is
\beq
    H_{\mu\nu}(q,t) =2 ( q_\mu q_\nu - g_{\mu\nu} Q^2 - t_\mu t_\nu ) \, ,
    \label{DY1}
\eeq 
with $t = k-k'$ 
the difference of the four momenta of the two incident quarks, $k$ and $k'$.

   In a heavy-ion collision, the  anisotropy in the Drell-Yan process arises only from the distributions of the initial quarks and antiquarks.   The imaginary part of the lowest-order photon polarization tensor in a heavy-ion collision  is    
   \beq
   \frac{1}{2}  \rho^{\mu\nu}(q) = (q_\mu q_\nu - g_{\mu\nu}Q^2)\langle 1 \rangle -\langle t_\mu t_\nu\rangle \, ,
         \label{pidy1}
\eeq
where  
     \beq
    \langle X \rangle &= & N_c
      \sum_f \frac{e_{\rm f}^2}{4\pi^2} \int {d^3\bar{k}} {d^3\bar{k}'} \ X   \delta^{(4)}(q-k-k')    f_{\vec k\,}   \bar f_{\vec k\,'}   \, ,  \nonumber \\
                  \label{pidy}
     \eeq
with the sum being over flavors, $e_{\rm f}$ denoting the quark charge; $+2/3$ for $u$-quarks, $-1/3$ for $d$-quarks, etc., and $N_c=3$ is the number of colors.  The generally anisotropic quark and antiquark 
distributions are denoted by  $f$ and $\bar f$, respectively.\footnote{The photon polarization operator is not that for an equilibrium system at finite temperature, owing to the fact that the electromagnetic sector in a heavy-ion collision never has adequate time to come into thermal equilibrium. The inverse processes in which 
a dilepton pair is absorbed, would lead to the distributions entering as $f_{\vec k}\bar f_{\vec k'} + (1-f_{\vec k} )(1- \bar f_{\vec k'}) = 1- f_{\vec k} -\bar f_{\vec k'}$ in the thermal equilibrium photon spectral function in full thermal equilibrium.}

  In this notation, the coefficients in the spectral distribution function are obtained by comparing Eqs.~(\ref{impilong}) and (\ref{pidy1}):
  \beq
   \rho_1^{\rm T} + \sin^2 \theta_q\, \rho_n =  2Q^2\langle 1\rangle - 2\langle (\varepsilon_1 t)^2\rangle \, , \nonumber\\
   \rho_2^{\rm T} = 2Q^2\langle 1\rangle - 2\langle (\varepsilon_2 t)^2\rangle \, ,\nonumber \\
   \rho^{\rm L} + ({\bar q}^0)^2\cos^2\theta_q\, \rho_n =  2Q^2\langle 1\rangle - 2\langle (\varepsilon_{\rm L} t)^2\rangle \, , 
\label{eq:rho_12Ln}
\eeq
with
\beq
      \rho_n = -\frac{4}{\bar{q}^0\sin 2 \theta_q} \langle  (\varepsilon_1 t)(\varepsilon_{\rm L} t)\rangle \, . 
      \label{rhontt}
\eeq
When the product of the distribution functions is invariant under parity as well as $\hat n\to -\hat n$, $\rho_n$ must vanish, as argued
after Eq.~(\ref{eq:trace-rho}).
 Explicitly, the orthogonality $(qt)=0$ implies $(\varepsilon_{\rm L} t) = -t^0/|\vec {\tilde q}\,|$, so that in Eq.~(\ref{rhontt}),  
$\langle \varepsilon_1 t)(\varepsilon_{\rm L} t)\rangle \sim \vec \varepsilon_1\cdot \langle t^0\vec t\,\,\rangle$.  But $\langle t^0\vec t\,\, \rangle$ can at most be proportional to a linear combination of $\vec q$ and $\hat n$, and  if $\hat n\to -\hat n$ is an invariance, the $\hat n$ term must vanish; then since $\vec q\cdot \vec \varepsilon_1 = 0$ one has $\rho_n = 0$.  
 In this case, Eq.~(\ref{eq:rho_12Ln}) implies 
\beq   
  \delta \rho^{\rm T} = - \langle (\vec \varepsilon_1\cdot \vec t\,)^2-(\vec\varepsilon_2\cdot \vec t\,)^2\rangle 
   = - \frac{3}{2}\langle t_z^2 - \vec t\,\,^2/3\rangle\sin^2\theta_q, \nonumber 
   \\ 
\eeq  
   which has the structure of a second spherical harmonic, and will thus select out the second spherical harmonic anisotropy in the particle distribution functions.  

   A simple approach to describing the anisotropy of the distributions is to assume an angular dependent temperature, so that, e.g., the
quark distribution of massless quarks becomes (with the chemical potential suppressed)
\beq
   f_{\vec k\,}  = \frac{1}{e^{\beta(\hat k)k}+1} \, .
  \label{anisof}
\eeq      
The parametrization of the angular dependent inverse temperature given in Ref.~\cite{RS} takes the form $\beta(\hat k) = \beta_0(1+\xi\cos^2\theta_k)^{1/2}$ where  $\theta_k$ is the angle between the quark momentum $\vec k$ and the anisotropy ($z$) axis.  The full calculation of the Drell-Yan dileptons in an ultrarelativistic heavy-ion collisions with such an anisotropic temperature has been given by Strickland et al.~\cite{ms-dy1,ms-dy2,ms-dy3,Ryblewski:2015hea}.

To illustrate, we expand the quark distribution functions, assumed to be of the form Eq.~(\ref{anisof}), to lowest order in the angular dependence of the temperature, writing 
\beq
  f_{\vec k\,} \simeq f_k^0 -  f_k^0 (1 - f_k^0)(\beta(\hat k)-\beta_0) k \, ,
  \eeq 
where $f_k^0$ is the distribution with $\beta(\hat k) = \beta_0$.  We see then that for weak anisotropy,
\beq
  \delta \rho^{\rm T} \sim  -\frac32 \langle t_z^2 -\vec t\,^2/3 \rangle \sim - {\beta_2} \eeq 
where $\beta_2 = \frac12 \int_{-1}^1 d(\cos\theta) \beta(\hat k) 
P_2(\cos\theta)=\beta_0 \xi/15 $ is the second spherical harmonic component of the temperature.   

 The terms in the photon spectral function $\rho_{\mu\nu}$  beyond the lowest order  Drell-Yan contribution are
found from the imaginary part of the polarization diagram  in Fig.~\ref{bubbles} with hard thermal loop corrections
and distribution anisotropies \cite{future}.

In practice, in order to obtain the dilepton spectra, one should integrate the spectral function, as given by Eq.~(\ref{pidy}), over the space-time volume of the collision vlume with an underlying model for the dynamics of $\xi(x)$ and $\beta_0(x)$ such as viscous \cite{HeinzJeon} or anisotropic hydrodynamics \cite{Strickland:2014pga}.  Previous work has shown that the high-energy dilepton rate computed in this manner is sensitive to the assumed initial momentum-space anisotropy of the plasma \cite{Ryblewski:2015hea} and that the momentum-space anisotropy of the quark-gluon plasma induces suppression of forward dilepton production \cite{ms-dy3}.

\section{Outlook}

   In this paper we have formulated the general structure of the spectral functions that describe the rate of virtual photon (dilepton) production in a heavy ion collision that is locally anisotropic in momentum space.  As we have demonstrated, momentum-space anisotropy induces new angular dependence in the transverse structure functions and, in the case of a non-parity symmetric momentum-space anisotropy, a new structure function $\rho_n$ appears.  As an example, we delineated the formalism for the leading-order Drell-Yan process.  The structure derived is not limited simply to production of virtual photons from a quark-gluon plasma, but encompasses all virtual photon production processes in collisions.

  With this full framework in place for relating polarization information in dilepton production to the underlying physical mechanisms, the next step will be to generalize prior calculations for real photon production in an anisotropic quark-gluon plasma \cite{Schenke:2006yp,Bhattacharya:2015ada}; also \cite{BH}.
A forthcoming publication will present a detailed calculation of the polarization of dileptons to order $\alpha_s$, the strong interaction fine structure constant, including Compton scattering of a gluon on a quark or antiquark to a virtual photon, and annihilation of a quark-antiquark pair into a gluon and virtual photon
\cite{future}.  In the computation of the spectral functions, we include space and time dependent anisotropic quark, antiquark, and gluon distributions, hard thermal loops, and soft scale processes, with the space-time evolution described by full 3+1 dimensional anisotropic hydrodynamics.

\section*{Acknowledgments}
   We are grateful to Dr.~Y.~Akiba for very helpful conversations about dilepton production; to Prof.~H.~Specht for discussions of the NA60 experiment,  to Dr.~B.~Friman and his group for detailed comparisons of his group's work and ours,  and to the ECT* in Trento, which hosted the workshop on ``New perspectives on Photons and Dileptons in Ultrarelativistic Heavy-Ion Collisions at RHIC and LHC" where we had valuable discussions.       We thank the RIKEN iTHES Project and iTHEMS Program for partial support during the course of this work.   This research was also supported in part by NSF Grant PHY1305891, JSPS Grants-in-Aid No.~25287066, and DOE Grant No. DE-SC0013470.  We thank the Aspen Center for Physics, supported in part by NSF Grant PHY1066292, where this paper was partially written.

\end{document}